# Multimodal CARS and SHG microscopy for label-free detection of collagen produced by hDFs in fibrin gel

**Leonardo Mortati[*], Carla Divieto,**
**and Maria Paola Sassi**

*INRIM – Istituto Nazionale di Ricerca Metrologica, Strada delle Cacce 91,Torino, 10135, Italy*
[*]*l.mortati@inrim.it*

**Abstract:** Label-free combined CARS and SHG microscopy techniques are used as powerful tool to follow the cells behavior in cell-scaffold construct for regeneration of tissues. Imaging of histological section of hDFs seeded in fibrin gel scaffold and imaging of collagen produced by hDFs in a time course experiment at different culture days (0, 7, 21, 42) is performed. A study on the limit of collagen detection of the imaging system is reported using sample prepared with different collagen concentrations. The results show that also the small amount of collagen produced by hDFs after few hours of incubation in fibrin gel is detected. Co-localization of hDFs and collagen is also reported in function of the culture days.

## 1. Introduction

Label-free microscopy techniques based on nonlinear optical process like Coherent Anti-Stokes Raman Scattering (CARS) and Second Harmonic Generation (SHG) are gaining a wide interest in biology imaging [1], due to their inherent deep tissue penetration [2,3], high 3D spatial resolution [4] and non-invasive chemical selection of target species preserving the samples from modifications induced by staining procedures.

Fluorescent staining drawbacks like bioconjugation efficiency and photobleaching in long term imaging can be completely avoided with label free techniques that usually have near-infrared lasers as sources, preventing also sample damaging and reducing light scattering in turbid media.

CARS microscopy, based on a four-wave mixing process, provides chemical contrast from resonant Raman-active molecular vibrations. CARS signal is generated when a pump laser and a Stokes laser beam with frequencies $\omega_p$ and $\omega_S$ respectively, have a frequency difference $\omega_p - \omega_S$ that coincides with a Raman-active resonant molecular vibration of the sample and the beams propagate respecting the phase-matching conditions. The generated CARS signal has a frequency of $2\omega_p - \omega_S$.

Lipid droplets and cell membranes of living fibroblasts have been imaged using CARS microscopy [5-6] looking for the $CH_2$ symmetric stretching vibration wavenumber at around 2845 $cm^{-1}$.

SHG microscopy technique is based on second order nonlinear optical process and it allows imaging contrast for non-centrosymmetric molecular ordered structures such as collagen [7-12], microtubule arrays [13], skeletal muscle myosin [14,15], cell membranes [16], and also cellulose [17].

Recently, it has been demonstrated that is possible to combine CARS and SHG techniques enabling multiple selective chemical contrasts [18-21].

In this work, the excitation wavelengths were selected in order to minimize Two-Photon Excitation Fluorescence (TPEF) from cell native fluorescent compounds.

Tissue engineering and regenerative medicine are young research fields with the main goal to replace or repair damaged tissues and organs. Many tissues are their targets, but the most established application field is the skin regeneration. Skin burns and diseases are the most common cause of skin damages and artificially produced dermis has been successfully implanted to replace the damaged tissue since the 1980s [22].

Type I collagen is the most abundant protein of skin providing, together with elastin, elastic properties, and with glycosaminoglycans, mechanical properties and is widely used as scaffold [23,24]. Collagen is synthesized by fibroblasts and in *in vitro* tests is a biomarker of fibroblasts functionality [25,26]. Autologous skin grafting is the main goal of the recent clinical research with the use of fibroblasts from the patient, seeded into a scaffold of fibrin gel and the implantation into the patient of the cells-scaffold construct [26]. The use of non-invasive techniques can allow a early detection of collagen production by human fibroblasts into scaffolds of fibrin gel in order to investigate this biological and natural process as marker of cell-scaffold interaction to translate the understandings into the clinic, extremely interested in the use of scaffold and autologous cells for tissues regeneration.

## 2. Methods

The laser sources for CARS and SHG microscopy are based on a passively mode-locked Nd:YVO$_4$ laser (Picotrain, HighQ Laser) emitting a continuous train of 10 ps pulses with a repetition rate of 76 MHz at 1064 nm. This laser is equipped with a SHG unit that duplicates with high efficiency the frequency, using a nonlinear Lithium-Triborate (LBO, $LiB_3O_5$) crystal. The 532 nm output of the SHG unit, has a pulse width of about 5 ps and it is used to synchronously pump an Optical Parametric Oscillator (OPO) (Levante Emerald, APE Berlin).

The OPO uses a nonlinear LBO crystal as a parametric amplifier in a resonant optical cavity. Wavelength tuning is achieved by changing the temperature of operation of the LBO crystal. The thermally induced change in refractive index changes the phase matching condition and consequently the two emission wavelengths of the OPO. The tuning range from 700 nm to 1020 nm for the signal wave and from 1110 nm to 2200 nm for the idler wave. Signal and idler beams exit collinear in our experiments, entering a Scanning Unit FluoView FV300 combined with the upright microscope Olympus BX51WI. The Z depth scanning is achieved by moving the focusing objective with a stepping motor.

In order to focus the excitation beams on the samples it was used an oil immersion objective (Olympus UPLASAPO 60XO; NA=1.35; W.D.=0.15 mm), fully compensate for both spherical and chromatic aberrations in the UV-VIS-NIR region. Microscope was modified to allow forward de-scanned detection of CARS and SHG signals that are collected through an Olympus UPLSAPO 20x objective; NA=0.75; W.D.=0.6 mm and focused on a PMT (model R3896, Hamamatsu) with a plane-convex lens with a focal length of 25 mm.

Cell membranes and rich lipidic structures were imaged using CARS looking for the $CH_2$ symmetric stretch Raman modes around 2844 $cm^{-1}$, tuning the pump and the Stokes beams to 924.1 nm and 1253.7 nm respectively and generating the CARS signal at a wavelength around 731.8 nm.

Collagen structures were imaged using SHG technique, tuning the OPO signal at 950 nm and detecting the corresponding halved wavelength at 475 nm.

Bandpass filters centered at 716 nm with 43-nm bandwidth (FF01-716/43, Semrock), 480 nm with 20-nm bandwidth (BP470-490, Chroma Technology) all coupled with shortpass filters with 770nm of cut-off wavelength (FF01-770/SP, Semrock) are used before the detector to further block the residual excitation beams and transmit the CARS and the SHG signals respectively.

CARS and SHG imaging are obtained on the same sample. The two measurements were performed in sequence, SHG images at low collagen concentrations were enhanced using *ImageJ* software.

The limit of collagen detection of the optical system was investigated using a set of five histological sections of fibrin gels spiked with various concentrations of rat tail collagen (1 mg/ml, 0.1 mg/ml, 0.05 mg/ml, 0.01 mg/ml and fibrin gel only) prepared by LGC laboratories (Teddington, UK), by following the briefly described protocol.

The fibrin gels were made by mixing in a 8 well plate thrombin (25 μg/mL) with neutralized collagen (1 mg/mL) and finally adding fibrinogen (5 mg/mL). The solution was then allowed to set at room temperature, then the gels were dehydrated in a graded series of ethanol washes and embedded in a low-melting paraffin wax to be sectioned (sliced of 10 μm and 30 μm) by the use of a microtome. Finally the wax was removed and the samples were re-hydrated.

Detection of collagen production by human dermal fibroblasts (hDFs) was investigated in a time-course experiment using several histological sections of cells fixed at different days in culture (day 0, 7, 21, 42).

Sections of fibrin gel containing hDFs producing collagen, were prepared by LGC as described: fibrin gels were made by mixing in a 8 well plate thrombin (25 μg/ mL) with cell growth medium containing hDFs ($1 \times 10^6$ cells in 315 μl of medium) and finally adding quickly fibrinogen (5 mg/mL). The solution was then allowed to set at room temperature and then the gels were fixed with 4% paraformaldayde (PFA). Then the gels were embedded in a low-melting paraffin wax and sectioned (sliced of 10 μm and 30 μm) with a microtome. Finally the wax was removed and the samples were re-hydrated.

## 3. Results and Discussion

The microscope limit of detection for collagen imaging was investigated analyzing the set of histological section of fibrin gel spiked with various concentration of rat-tail type I collagen.

Images of fibrin gel spiked with rat-tail type I collagen were obtained doing a single slice measurement with a XY pixel pitch ranging from 0.16 μm/pixel to 0.33 μm/pixel depending by the analyzed sample. For each experiment up to 7 images were averaged using an adaptive Kalman filtering.

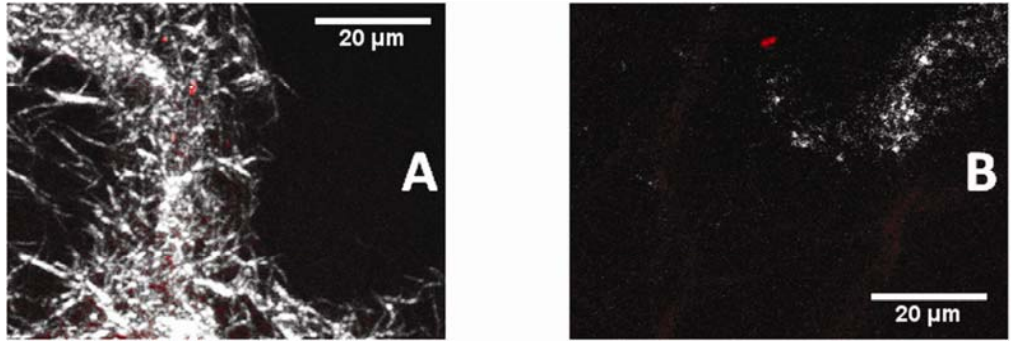

Fig. 1 Images of fibrin gel (in red) spiked with collagen (in white) at concentrations of 1 mg/ml (A) and of 0.1 mg/ml (B)

Images of collagen spiked in fibrin gel at concentration of 1 mg/ml and 0.1 mg/ml are showed respectively in Fig. 1A and 1B. Collagen (in white) is detected using SHG technique while fibrin gel (in red) is acquired using CARS generated at the vibrational resonance of 2844 $cm^{-1}$. At this concentration collagen tends to form complex structures along the fibrin gel.

At lower collagen concentrations the SHG signal is weak and an image processing is necessary to improve the signal contrast. SHG signal strength of acquired images is improved using *ImageJ* software with a contrast enhancement (0.4% of saturate pixels with histogram equalization) followed by a Gaussian blurring (one pixel size) and a manual adjustment of the brightness/contrast levels.

Using this method it is possible to localize also low concentrations of collagen, obtaining a detection limit beyond 0.01 mg/ml, the lowest collagen concentration spiked in the fibrin gel. In Fig. 2A and Fig. 2B are shown the superimposed images of collagen traces (in white) detected using SHG technique and of the fibrin gel structure (in red) detected using CARS, respectively at collagen concentration of 0.05 mg/ml and 0.01 mg/ml. At low concentrations collagen tends to agglomerate in small spots of few micrometer size. Fig. 2C shows the CARS and the SHG superimposed images of a sample with only fibrin gel, where collagen was not spiked. No relevant SHG signal was detected, confirming the absence of collagen in the sample. This confirmation is also taken into account as a negative control for SHG microscopy of the collagen (absence of false positive detection).

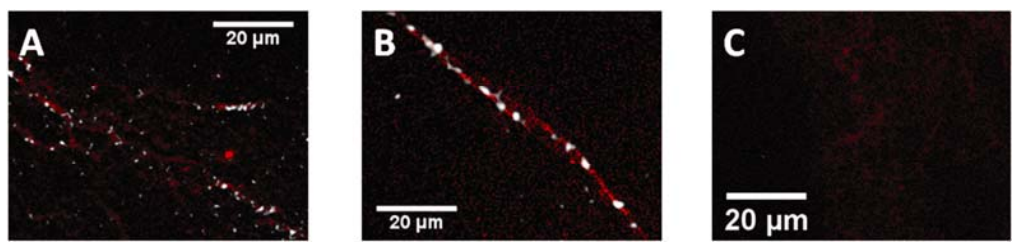

Fig. 2 Images of fibrin gel (in red) spiked with collagen (in white) at concentrations of 0.05 mg/ml (A) and of 0.01 mg/ml (B) and without collagen (C).

Collagen produced by hDFs was detected and localized in a time-course experiment using several histological sections of cells fixed at different days in culture (day 0, 7, 21, 42).

Three-dimensional imaging was performed slicing the sample with Z-axis steps of about 800 nm between two adjacent slices. The 3D extension of the cells in the fibrin gel determined the overall Z scan range. The XY pixel pitch was equal to 0.16 μm/pixel for the measurement related to the day 0 and equal to 0.33 μm/pixel for all the other measurements. The acquired image size was adapted to cells shape in each experiment.

Each slice was obtained adaptively averaging with a Kalman filter up to seven images and the 3D imaging lasted about 5 minutes according to the Z scan range.

Pixel dwell time was around 10 μs and the average power at the sample was about 25 mW for the pump signal and less than 10 mW for Stokes signal.

Cells were localized using CARS microscopy looking for cells rich lipidic structures in correspondence of the $CH_2$ symmetric stretch at 2844 $cm^{-1}$.

Collagen detection was done using SHG microscopy technique doubling the imaging experiment and keeping the same dimensional and temporal parameters of the related CARS imaging and average excitation power at the sample was about 20 mW.

At the end of the experiment a maximum intensity Z-projection image was extracted from the obtained slices, creating a single picture with all the interesting extension of the occupied cells volume.

As it is showed in Fig. 3, at day 0 there is not an important amount of collagen (in white) produced by hDFs (in red). However, some small collagen traces are already visible indicating the powerful of the method in revealing also a slight quantity of collagen produced during the first hours of cells incubation inside the fibrin gel.

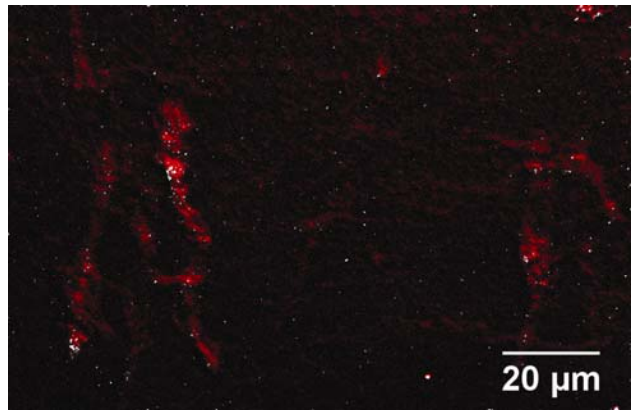

Fig. 3 Image of hDFs (in red) and collagen produced (in white) at culture day 0.

Confirmed by the negative control result, the poor spreading of the SHG signal is with high consistency generated by collagen and native cell autofluorescence can be excluded using 950 nm as excitation wavelength for SHG.

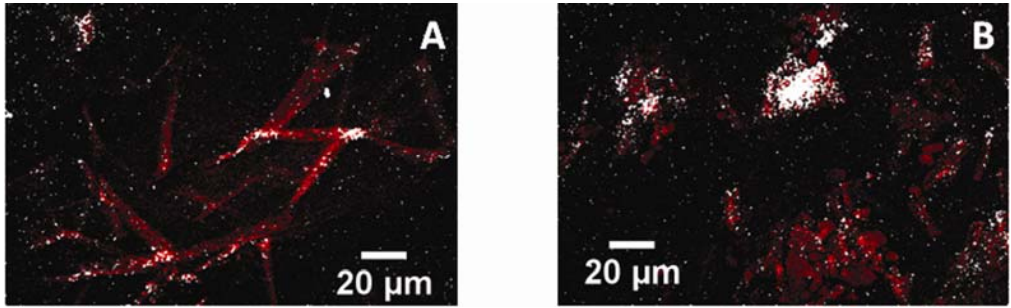

Fig. 4 Image of hDFs (in red) and collagen produced (in white) at culture day 7 (A) and culture day 21 (B).

At day 7 and day 21 as it can be seen in Fig. 4A and Fig. 4B respectively, hDFs (in red) produced a clearly localizable amount of collagen (in white). At day 21 collagen seems to be well organized on the whole surface of a group of cells.

In Fig. 5 is showed the image acquired with the hDFs at day 42. The cells (in red) produced an high amount of collagen (in white) that cover widely most of the cells surface.
In order to quantify the amount of collagen produced by the hDFs in culture and, at the same time, to evaluate the spatial localization of collagen in respect to the cell location, the co-localization percentage of the collagen surface over the cells surface was analyzed. In this way it is possible to study the production of collagen in terms of spatial and temporal increase of its amount.

A specific *ImageJ* plug-in, written in Java language, was developed in order to analyze the co-localization percentages of all the samples investigated at different culture days, starting from binary images obtained using a manual threshold of hDFs (CARS) and collagen (SHG) acquired images.

The co-localization algorithm creates a new binary image (Fig. 6C) in which only the common surface belonging to both input images (Fig. 6A and 6B) are visualized.

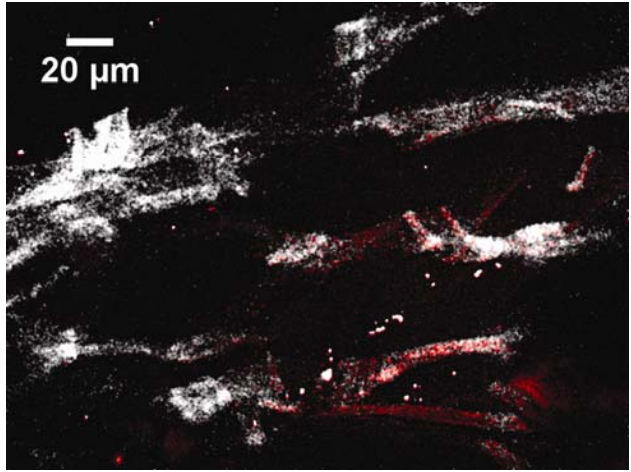

Fig. 5 Image of hDFs (in red) and collagen produced (in white) at culture day 42.

Then the percentage of the common surface (Fig 6C) with respect to the total cells surface referred to the CARS binary image (Fig 6A) is computed.

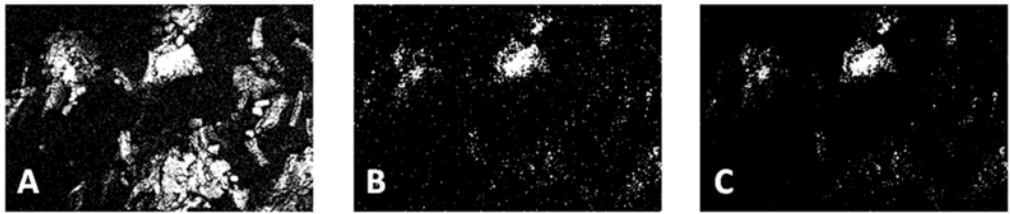

Fig. 6 Binary images of hDFs acquired with CARS technique (A), of collagen produced by hDFs acquired with SHG technique (B) and the colocalization image of hDFs and collagen (C).

As it could be expected, the co-localization percentage increased with the culture time (Fig. 7), indicating that collagen produced by the hDFs increased with time.

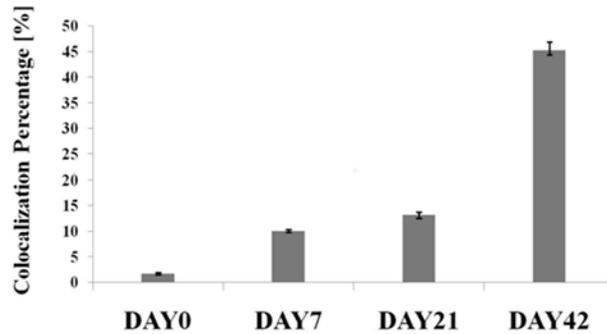

Fig. 7 Co-localization percentage of the collagen surface referred to hDFs surface at different culture days.

## 4. Conclusion

This work demonstrates that CARS - SHG combined technique is a powerful tool to follow label free cells behavior, in this case the collagen production, by detecting and localizing in a 3D fibrin gel matrix both cells (i.e. hDFs) and cell-produced collagen.

The microscopy limit of detection for collagen has been investigated in the fibrin gel with spiked collagen at different concentrations and a limit down to 0.01 mg/ml was proved.

This low detection limit allowed to test the cells production of collagen in a very early cell culture (less than 24 hours).

In order to better understand the amount of collagen produced in relation with the cells surface, an analysis was conducted estimating the co-localization of the SHG (collagen) signal and the CARS (cells) signal with a specifically developed *imageJ* plug-in. With this method it was possible to determine that collagen co-localization percentage tends to increase with the culture time showing that at day 42 about the half of the entire cell surface was covered with collagen.

## Acknowledgements

The authors like to thanks Gary Morley from LGC (UK) for providing the samples used in this work. This work was partially supported by ERA-NET Plus Project ReGenMed – Grant Agreement No 217257 and by Regione Piemonte CIPE 2007 Converging technology Project Metregen - Grant Agreement 20/07/2007.